\documentclass[aps,prl,twocolumn,groupedaddress]{revtex4}
\usepackage{amssymb}
\usepackage{amsmath}

\input{tcilatex}

\begin{document}

\title{ The absorber hypothesis of electrodynamics}
\author{Jayme De Luca}
\email[author's email address:]{ deluca@df.ufscar.br}
\affiliation{Universidade Federal de S\~{a}o Carlos, \\
Departamento de F\'{\i}sica\\
Rodovia Washington Luis, km 235\\
Caixa Postal 676, S\~{a}o Carlos, S\~{a}o Paulo 13565-905\\
Brazil\\
}
\date{\today }

\begin{abstract}
We test the absorber hypothesis of the action-at-a-distance electrodynamics
for globally-bounded solutions of a finite-particle universe. We find that
the absorber hypothesis forbids globally-bounded motions for a universe
containing only two charged particles, otherwise the condition alone does
not forbid globally-bounded motions. We discuss the implication of our
results for the various forms of electrodynamics of point charges.
\end{abstract}

\maketitle

\section{Introduction}

Action-at-a-distance electrodynamics was formulated in 1945 \cite{Fey-Whe},
which is relatively late in electromagnetic history. The original motivation
was the regularity of the point-charge limit, but the theory has so many
desirable physical properties that it is surprising it was not considered
earlier for the other reasons. In the original articles, Wheeler and Feynman 
\cite{Fey-Whe} further promoted the absorber hypothesis, whereby
action-at-a-distance electrodynamics reduces to Dirac%
\'{}%
s electrodynamics of point charges with retarded-only fields \cite{Dirac}.
The absorber hypothesis\cite{Fey-Whe} states that the universe absorbs all
future and past radiation, i.e., that the universal far-fields vanish at all
times. If true, the absorber hypothesis would be useful to approximate the
electrodynamics of the other charges of a large bounded universe, thereby
avoiding a many-body problem, but the approximation has never been tested.
Here we show that the absorber hypothesis fails for a universe consisting of
two charges with globally-bounded interparticle distances. Otherwise, for
universes with three- or more- point charges the absorber hypothesis alone
is no obstacle for globally-bounded motions. For these possible universal
motions it remains to be understood if the absorber hypothesis is an
additional property that holds for a special class of globally-bounded
orbits, like neutrally-stable orbits for example, or just true for any
globally bounded orbit of a universe containing a large enough number of
charges.

The absorber hypothesis, henceforth denoted A.H., \cite{Fey-Whe} is most
easily expressed in terms of the advanced and retarded electromagnetic field
tensor of each particle, respectively $F^{(i)adv}$ and $F^{(i)ret}$ , as 
\begin{equation}
\dsum\limits_{i}[F^{(i)ret}+F^{(i)adv}]=o(\frac{1}{r})  \label{absorber}
\end{equation}%
when $r\rightarrow \infty $, as discussed in \cite{Narlikar}. The fall
faster than $(1/r)$ defined by Eq. (\ref{absorber}) implies the vanishing of
the semi-sum of the radiation fields, which is called perfect absorption. As
argued by Wheeler and Feynman \cite{Fey-Whe}, Eq. (\ref{absorber}) includes
a \ combination of incoming and outgoing waves, and the only way to achieve
a cancellation at all times is if both the retarded sum and the advanced sum
vanish separately, i.e.,

\begin{eqnarray}
\dsum\limits_{i}F^{(i)ret} &=&o(\frac{1}{r}),  \label{absorber1} \\
\dsum\limits_{i}F^{(i)adv} &=&o(\frac{1}{r}),  \label{absorber2}
\end{eqnarray}%
when $r\rightarrow \infty $. In the sequel of \ their work of 1945, Wheeler
and Feynman used Eqs. (\ref{absorber1}) and (\ref{absorber2}) to establish
the important condition

\begin{equation}
\dsum\limits_{i}[F^{(i)ret}-F^{(i)adv}]=o(\frac{1}{r}),  \label{dif}
\end{equation}%
for the source-free retarded-minus-advanced field, which implies its
vanishing everywhere by the Cauchy problem.\ The various electrodynamics of
point charges originated from Maxwell%
\'{}%
s electrodynamics, by a procedure inaugurated by Lorentz\cite{Dirac,
Nodvik,EliezerReview}. Maxwell's equations are time-reversible and therefore
the general solution is given by a linear combination of the retarded Green
function and the advanced Green function\cite{Jackson}, each combination
generating a different electrodynamics of point charges\cite{EliezerReview}.
In the general case the far-electric field of a point charge involves an
arbitrary parameter $\chi $ \cite{Jackson}, 
\begin{equation}
\boldsymbol{E}=\frac{1}{2}(1-\chi )\mathbf{E}^{adv}+\frac{1}{2}(1+\chi )%
\mathbf{E}^{ret},  \label{genE}
\end{equation}%
while the far-magnetic field is given by%
\begin{equation}
\boldsymbol{B}=\frac{1}{2}(1-\chi )\boldsymbol{n}^{+}\times \mathbf{E}^{adv}-%
\frac{1}{2}(1+\chi )\boldsymbol{n}^{-}\times \mathbf{E}^{ret},  \label{genB}
\end{equation}%
where unit vectors $\boldsymbol{n}^{\pm }$ point away from the
advanced/retarded position of the charge, respectively \cite{Jackson} and in
our unit system the speed of light is $c=1$. For a spatially bounded
universe we can take a sphere of radius much larger than the universal
radius, such that $\boldsymbol{n}^{+}=\boldsymbol{n}^{-}\equiv \boldsymbol{n}
$, and the Poynting vector $\boldsymbol{P}=\boldsymbol{E}\times \boldsymbol{B%
}$ evaluated with Eqs. (\ref{genE}) and (\ref{genB}) is 
\begin{equation}
\boldsymbol{P=}\frac{1}{4}\{(1-\chi )^{2}|\mathbf{E}^{adv}|^{2}-(1+\chi
)^{2}|\mathbf{E}^{ret}|^{2}\}\boldsymbol{n}  \label{PoyntingX}
\end{equation}%
where single bars denote Euclidean modulus, and we have used the
transversality of the far-fields, $\boldsymbol{n\cdot }\mathbf{E}^{ret}=%
\boldsymbol{n\cdot }\mathbf{E}^{adv}=0$.

Notice that the A.H. is a stronger condition than the vanishing of the
Poynting flux of the semi-sum field obtained from Eq. (\ref{PoyntingX}) with 
$\chi =0$. For instance, the A.H. \emph{implies} the vanishing of the flux
Eq. (\ref{PoyntingX}) for any $\chi $ because $|\mathbf{E}^{ret}|^{2}=|%
\mathbf{E}^{adv}|^{2}=0$ by Eqs. (\ref{absorber1}) and (\ref{absorber2}),
but not vice-versa. For example the circular two-body orbit \cite%
{Schoenberg,Schild} of the action-at-a-distance theory has a vanishing
angular average of $\boldsymbol{P}$ but it does not satisfy the A.H.
(neither Eq. (\ref{absorber1}) nor Eq. (\ref{absorber2}) hold). Therefore,
studying orbits with a vanishing Poynting flux of retarded fields is
relevant for two different electromagnetic theories of point charges: (i) In
the action-at-a-distance electrodynamics it is a necessary condition for
globally bounded two-body orbits \emph{satisfying }the A.H., because the
A.H. implies Eq. (\ref{absorber1}), and (ii) In the Dirac theory with
retarded-only fields\cite{Dirac} it is a necessary condition for globally
bounded orbits in general, otherwise there are energy losses to infinity. In
the following we investigate the implications of condition (\ref{absorber1})
for a globally bounded orbit, regardless of the equations of motion of each
electrodynamics of point charges. The paper is divided as follows: In
Section 1 we consider the consequences of vanishing the flux of the retarded
fields for a universe containing a single point charge and we find that the
A.H. is always satisfied. In Section 2 we consider the consequences of
vanishing the energy flux of the retarded fields for a globally bounded
orbit of a universe with two charges, and we find that condition (\ref%
{absorber1}) forbids globally bounded motions. For universes with three- or
more charges, condition (\ref{absorber1}) alone no longer forbids globally
bounded motions. Last, in Section 3 we discuss the implications of the
results for the existing versions of electrodynamics of point charges and
verify the results on some known solutions of many-body motion.

\section{\protect\bigskip Universe with a single charge}

We henceforth adopt a unit system where the speed of light is $c=1$. To
apply conditions (\ref{absorber1}) and (\ref{absorber2}) to a spatially
bounded one-body orbit takes an inertial frame and a sphere of large radius $%
R$ centered at the origin. The space-time points $(t,R\mathbf{n})$ on the
sphere are specified by the time $t$ and the unit vector $\mathbf{n}$. As
far a necessary condition is concerned, it suffices to consider the
condition for the far-retarded fields, Eq. (\ref{absorber1}), the advanced
condition Eq. (\ref{absorber2}) representing exactly the same obstruction.
It is also sufficient to apply the condition to the retarded-far-electric
field $\mathbf{E}_{1}^{ret}(t,\mathbf{n})$ only, since the far-magnetic
field is proportional to $\mathbf{E}_{1}^{ret}(t,\mathbf{n})$ by%
\begin{equation}
\mathbf{B}_{1}^{ret}(t,\mathbf{n})=\mathbf{n}\times \mathbf{E}_{1}^{ret}.
\label{far-magnetic1}
\end{equation}%
The retarded far-electric field of a charge $q_{1}$ at a space-time point $%
(t,R\mathbf{n})$ is given by the Lienard-Wiechert formula\cite{Jackson}

\begin{equation}
\mathbf{E}_{1}^{ret}(t,\mathbf{n})=q\frac{\mathbf{n}\times \lbrack (\mathbf{%
n-\mathbf{v}}_{1}\mathbf{\mathbf{(}}t_{1}\mathbf{)}\times \mathbf{a}%
_{1}(t_{1})]}{(1-\mathbf{n\cdot v}(t_{1}))^{3}R}.  \label{far-electric1}
\end{equation}%
In Eq. (\ref{far-electric1}), unit vector $\mathbf{n}$ points from the
charge 's retarded position $(t_{1},\mathbf{x}_{1})$ to the space-time point 
$(t,R\mathbf{n})$ while$\mathbf{v}_{1}(t_{1})$ and $\mathbf{a}_{1}(t_{1})$
are respectively the Cartesian velocity and Cartesian acceleration of the
point charge in the past light-cone of the observation point $(t,R\mathbf{n}%
) $. The time of particle $1$ in light-cone with $(t,R\mathbf{n})$ is given
by%
\begin{equation}
t_{1}=t-|\mathbf{x}_{1}(t_{1})-R\mathbf{n|,}  \label{lightcone1}
\end{equation}%
where single bars stand for Cartesian distance. Equation (\ref{lightcone1})
is approximated at large values of $R$ by 
\begin{equation}
t_{1}\mathbf{=}t-R+\mathbf{n}\cdot \mathbf{x}_{1}(t_{1})\mathbf{.}
\label{cone1}
\end{equation}%
Equation (\ref{cone1}) defines $t_{1}$ as an implicit function of time $t$
with derivative%
\begin{equation}
\frac{dt_{1}}{dt}=\frac{1}{(1-\mathbf{n\cdot v}_{1}(t_{1}))}.  \label{dt1dt}
\end{equation}%
Using Eq. (\ref{dt1dt}), the far-field (\ref{far-electric1}) can be
expressed simply as 
\begin{equation}
\mathbf{E}_{1}^{ret}(t,\mathbf{n})=\frac{q_{1}}{R}\frac{d^{2}}{dt^{2}}[%
\mathbf{n}\times \mathbf{x}_{1}(t_{1})],  \label{farman1}
\end{equation}%
where $\mathbf{x}_{1}(t_{1})$ is the position of particle $1$ at time $t_{1}$%
. The retarded time $t_{1}$ is a function of time $t$ by Eq. (\ref%
{lightcone1}), and condition (\ref{absorber1}) reduces to the vanishing of
Eq. (\ref{farman1}). We recall that the A.H. condition (\ref{absorber1})
implies the vanishing of the Poynting vector of the retarded fields ( Eq. (%
\ref{PoyntingX}) with $\chi =1$), i.e.,%
\begin{equation*}
\mathbf{P}_{1}^{ret}=-|\mathbf{E}_{1}^{ret}|^{2}\mathbf{n}=0\mathbf{,}
\end{equation*}%
because the A.H. requires $\mathbf{E}_{1}^{ret}=0$. The A.H. is a simple
ordinary differential equation 
\begin{equation}
\frac{\mathbf{n}}{R}\times \frac{d^{2}}{dt^{2}}\mathbf{x}_{1}(t_{1})=0,
\label{vanishing1}
\end{equation}%
with general solution 
\begin{equation}
\mathbf{x}_{1}(t_{1})=\mathbf{D}_{1}(\mathbf{n})+\mathbf{n}f_{1}(t,\mathbf{n}%
),  \label{dipole1}
\end{equation}%
where $\mathbf{D}_{1}(\mathbf{n})$ is an arbitrary bounded function of $%
\mathbf{n}$ that can be assumed to satisfy $\mathbf{n}\cdot \mathbf{D}_{1}%
\mathbf{=0}$ and $f_{1}(t,\mathbf{n})$ is a $C^{2}$bounded function of time.
The derivative of Eq. (\ref{dipole1}) yields the velocity%
\begin{equation}
\mathbf{v}_{1}(t_{1})=(1-\mathbf{n}\cdot \mathbf{v}_{1}(t_{1}))\frac{%
\partial f_{1}}{\partial t}\mathbf{n}.  \label{velocity1}
\end{equation}%
According to Eq. (\ref{cone1}) one can vary $\mathbf{n}$ in a cone with axis
along $\mathbf{x}_{1}(t_{1})\neq 0$ while leaving $t_{1}$ and $t$ fixed so
that the left-hand side of Eq. (\ref{velocity1}) is fixed. This is seen to
be impossible because the right-hand side of Eq. (\ref{velocity1}) varies
unless%
\begin{equation}
\frac{\partial f_{1}}{\partial t}=0,
\end{equation}%
which must be the case. Therefore $\mathbf{v}_{1}(t_{1})=0$ and the particle
is resting at the origin of some inertial frame for the only consistent
solution. No surprises arise in this one-charge case, the A.H. requires only
that the charge is resting in some inertial frame. Since an isolated charge
suffers no force in the action-at-a-distance theory, it moves at a constant
velocity and we can always find an inertial frame where it is resting, so
that the A.H. is always fulfilled !

\bigskip

\section{Universe consisting of two charges}

Unlike the case of a single charge, where a trivial resting orbit \emph{is}
acceptable, a two-body bounded orbit with both charges resting a finite
distance apart is unacceptable because the Coulombian attraction would cause
an acceleration incompatible with constant rest. \ Assuming a globally
bounded motion exists, it should be inside a sphere of radius $\rho <<R$ .
The Poynting vector of the retarded fields on the surface of this sphere is 
\begin{equation}
\mathbf{P}^{ret}=-|\mathbf{E}_{1}^{ret}+\mathbf{E}_{2}^{ret}|^{2}\mathbf{n}.
\label{Poynting}
\end{equation}%
Here we consider only the case of opposite charges. Our next result is true
for the case of two arbitrary charges as well, but the mathematical details
are surprisingly more elaborate and since the protonic and the electronic
charges are known to be equal to an incredible precision\cite{Starus}, we
consider here only the equal-charge problem. Henceforth charge $1$ is
supposed positive and equal to $q$ while charge $2$ is negative and \ equal
to $-q$. Again the absorber condition (\ref{absorber1}) is equivalent to the
vanishing of the flux of the retarded fields, i.e., 
\begin{equation}
\mathbf{E}_{1}^{ret}+\mathbf{E}_{2}^{ret}=\frac{q\mathbf{n}}{R}\times \frac{%
d^{2}}{dt^{2}}(\mathbf{x}_{1}(t_{1})-\mathbf{x}_{2}(t_{2}))=0.
\label{vanishing}
\end{equation}%
The minus sign in Eq. (\ref{vanishing}) is because the charges are opposite.
Analogously to the time of particle $1$, in Eq. (\ref{vanishing}) the time
of particle $2$ is given by 
\begin{equation}
t_{2}\mathbf{=}t-R+\mathbf{n}\cdot \mathbf{x}_{2}(t_{2})\mathbf{.}
\label{cone2}
\end{equation}%
Notice also that Eqs. (\ref{cone1}) and \ (\ref{cone2}) yield an implicit
relation between $t_{1}$and $t_{2}$, 
\begin{equation}
t_{1}-t_{2}=\mathbf{n}\cdot (\mathbf{x}_{1}(t_{1})-\mathbf{x}_{2}(t_{2})).
\label{t1t2}
\end{equation}%
Equation (\ref{vanishing}) has a general bounded solution of type%
\begin{equation}
\mathbf{x}_{1}(t_{1})-\mathbf{x}_{2}(t_{2})=\mathbf{D}(\mathbf{n})+\mathbf{n}%
f(t,\mathbf{n})+t\boldsymbol{S}(\mathbf{n}),  \label{dipole}
\end{equation}%
where again $\mathbf{D}(\mathbf{n})$ is an arbitrary bounded function of $%
\mathbf{n}$ satisfying $\mathbf{n}\cdot \mathbf{D=0}$ , $f(t,\mathbf{n})$ is
a $C^{2}$bounded function of time and $\boldsymbol{S}(\mathbf{n})$ is a
bounded vector function of $\mathbf{n}$. It follows from Eqs. (\ref{t1t2})
and (\ref{dipole}) that $f(t,\mathbf{n})=(t_{1}-t+R)-(t_{2}-t+R)$ and the
condition of a globally-bounded orbit implies $\boldsymbol{S}(\mathbf{n})=0$%
. We therefore rewrite Eq. (\ref{dipole}) as 
\begin{equation}
\mathbf{x}_{1}(t_{1})-\mathbf{x}_{2}(t_{2})=\mathbf{D}(\mathbf{n}%
)+[(t_{1}-t)-(t_{2}-t)]\mathbf{n}.  \label{express}
\end{equation}%
The derivative of Eq. (\ref{express}) respect to time yields 
\begin{equation}
\frac{\mathbf{v}_{1}(t_{1})}{(1-\mathbf{n}\cdot \mathbf{v}_{1}(t_{1}))}-%
\frac{\mathbf{v}_{2}(t_{2})}{(1-\mathbf{n}\cdot \mathbf{v}_{2}(t_{2}))}%
=K_{12}\mathbf{n,}  \label{velocity}
\end{equation}%
where%
\begin{equation}
K_{12}=\frac{1}{(1-\mathbf{n\cdot v}_{1}(t_{1}))}-\frac{1}{(1-\mathbf{n\cdot
v}_{2}(t_{2}))}.  \label{defK}
\end{equation}%
For arbitrary $t_{1}$ and $t_{2}$ \ it is possible to move $\mathbf{n}$ in a
cone with axis along $\mathbf{x}_{1}(t_{1})-\mathbf{x}_{2}(t_{2})\neq 0$ in
a way that fixes $t_{1}$ and $t_{2}$. It is important to observe that the
time $t$ of the observation point also \emph{changes }along this variation,
as needed by Eqs. (\ref{cone1}) and (\ref{cone2}).\ Along this variation,
the unitary vector $\mathbf{n}$ describes a cone that can be at the best
tangent to the plane defined by the fixed vectors $\mathbf{v}_{1}(t_{1})$
and $\mathbf{v}_{2}(t_{2})$. \ Therefore the only possibility is that $%
K_{12}=0$, such that the velocity vectors on the left-hand side of Eq. (\ref%
{velocity}) must be collinear. It is further possible to show with Eq. (\ref%
{defK}) using $K_{12}=0$ and Eq. (\ref{velocity}) that 
\begin{equation}
\mathbf{v}_{1}(t_{1})=\mathbf{v}_{2}(t_{2}),  \label{rest12}
\end{equation}%
which can be used to prove that each velocity is piecewise constant, as
follows. Since Eq. (\ref{rest12}) is valid for arbitrary $t_{1}$ and $t_{2}$
satisfying the light-cone condition (\ref{t1t2}), one can fix $t_{1}$ while
moving $t_{2}$ to a maximal interval by playing with $t$ and $\mathbf{n}$,
such that the velocity is constant in the maximal interval determined by Eq.
(\ref{t1t2}). It is interesting to notice that the velocity must remain
constant for the maximal time equal to the interparticle separation
predicted by Eq. (\ref{t1t2}), i.e.,

\begin{equation}
a=|\mathbf{x}_{1}(t_{1})-\mathbf{x}_{2}(t_{2})|.  \label{dist}
\end{equation}%
The distance $a$ is itself constant while the particles have the same
velocity, as of Eq. (\ref{rest12}). If the particles never collide, the
particle separation $|\mathbf{x}_{1}(t_{1})-\mathbf{x}_{2}(t_{2})|$ must be
bounded from below, and the only physical motion having piecewise-constant
velocity and piecewise-constant separation on bounded intervals as derived
above would be motion of both particles with the same constant velocity.
This motion would be impossible because the Coulombian force from the other
particle at a finite distance would necessarily produce acceleration. The
remaining options left would be spiky orbits with a discontinuos velocity at
constant particle separation, where the two particles jump together, and it
would be unphysical. Our conclusion is then that for two isolated charges
there is no globally-bounded $C^{2}$ orbit satisfying the A.H. Our analysis
naturally came down to collinear collision orbits, which have zero angular
momentum like the quantum ground-state of the hydrogen atom. In the
action-at-a-distance theory these collision orbits \emph{are} possible and
have been calculated in Ref. \cite{EFY2} and could be the candidates to
satisfy the A.H. at least marginally (for example, with minimal radiative
losses over a very large time). Unfortunately such collisions terminates
after a finite time\cite{EFY2} and form a composite particle which does not
radiate because it moves as one charge, so that again one does not have two
charges in non-trivial bounded motion. For the other forms of
electrodynamics the very existence of collision orbits is
problematic;---There is yet no result available for two-body motion with
arbitrary masses in the Lorentz-Dirac theory, and surprisingly the only
existing result used an infinitely-massive second particle and concluded
that no collision solution exists\cite{Eliezertheorem}. Last, for three- or
more particle systems no incompatibility is found as far as the A.H. is
concerned. In the Lorentz-Dirac theory it would be necessary that a
globally-bounded solution did not radiate energy so that our result suggests
that such solution does not exist for two-charge universes. It remains to be
studied if either a third distant charge can already support a
globally-bounded physical orbit in the Dirac theory or if a large number of
charges is needed. The above result does not necessarily put an end to the
interest in the Lorentz-Dirac theory, but it suggests the supporting
universe should contain more-than-two charges in globally-bounded motion,
which seems to be the case of our universe.

\section{\protect\bigskip Discussions and Conclusion}

We have seen that the A.H. forbids globally bounded orbits for a universe
with only two charges. For universes with three- or more charges the A.H.
condition alone does not exclude globally-bounded electromagnetic orbits.
Moreover, since Eq. (\ref{velocity}) does not involve interparticle
distances, the globally-bounded non-radiating orbit could even be such that
the third charge is significantly separated from the other two. \ However,
since the A.H. is only a necessary condition, we can not conclude that
globally bounded orbits satisfying the A.H. do exist for
more-than-two-charges universes, and for that it would be necessary to deal
with the neutral-delay equations of motion, or to go by inspection of known
solutions\cite{Schoenberg,Schild}. The investigation of the possible
solutions of these neutral-delay equations is still an open problem.
Non-radiating motions of spatially-extended charge distributions with
retarded-only fields have already been studied \cite{radiationless}, but so
far none with only three point charges was reported in the literature.

The consequences of our results depend if one is using the Dirac theory with
retarded fields or the action-at-a-distance electrodynamics, as follows:
(i)\ For the Dirac theory our results strongly suggest that no
globally-bounded solution exists at all for the Lorentz-Dirac equation with
two charges, because such orbit would be leaking energy to infinity. This
physical limitation of the Dirac theory makes it impossible to consider the
other charges of a large universe as a perturbation for a globally-bounded
two-body orbit, but rather these other charges are an essential ingredient
for the bounded motion to exist. In Ref.\cite{stiff-hydrogen} the vanishing
of the flux of retarded fields, Eq. (\ref{absorber1}), was used as a
condition for long-lived orbits;-- If these orbits conjectured in Ref.\cite%
{stiff-hydrogen} are only long-lived and eventually ionize or decay, there
is no contradiction with our present results. Otherwise if these orbits are
to be globally-bounded, our present result requires at least a third charge
somewhere, in the best scenario for the Dirac theory. (ii) On the other
hand, for the action-at-a-distance electrodynamics bounded two-body orbits
do exist\cite{Schoenberg,Schild}, and it would be physically sensible to
consider the other charges of a large universe as a perturbation for a
globally-bounded two-body orbit. Notice that the Schoenberg-Schild orbits 
\cite{Schoenberg,Schild} do not satisfy the A.H. (\ref{absorber1}), in
agreement with our result that holds for any globally-bounded orbit. These
circular orbits do no leak energy to infinity because the total flux of the
semi-sum field vanishes, \emph{even though} the far-fields themselves do no
vanish and neither the flux of the retarded-fields vanishes. The natural
physical interpretation of condition (\ref{absorber1}) as used in Ref.\cite%
{stiff-hydrogen} is as follows;-- Let us promote the A.H. to a condition
that is respected by the dynamics of a stable universe with a large number
of charges. Along such dynamics, if a two-body circular orbit is formed, the
other universal charges must provide reaction far-fields to add linearly to
these and enforce the A.H., since no bounded two-body orbit can possibly
satisfy the A.H. alone. In this perspective, the globally-bounded two-body
orbit is disturbing the universe by disturbing the A.H. boundary condition
at infinity, and it would be physically desirable to minimize the strength
of the offending retarded far-fields of the two-body orbit, so that lesser
universal reaction is required to enforce the A.H. For that the mechanism
suggested in Ref. \cite{stiff-hydrogen}, i.e., interference with a beat of
the fast solenoidal motion of a deformed two-body orbit, would be a
physically desirable perturbation. It remains to be researched if such
solenoidal two--body orbits do exist as novel \ nontrivial solutions of
unperturbed two-body motion in the action-at-a-distance theory. In a large
bounded universe the far-fields of the other particles must be included in
the flux calculation, which is implemented in the random electrodynamics\cite%
{Boyer} in the form of a non-A.H. boundary condition for the far-fields. The
inclusion of the universal far-fields in some estimates for the physical
magnitudes of solenoidal two-body orbits was implemented using a model
called dissipative Fokker electrodynamics in Ref. \cite{dissipativeFokker}.

In has been common use in the derivations of point-charge electrodynamics
that a particular choice of Green function is enough to describe the most
general physical dynamics, and it is even desirable that it should be so for
simplicity. Of course one can not rule out that a particular choice leads to
equations with no solution of physical interest, or even worse, that the
Green function yielding the correct physics could be different for different
problems. Assuming a particular choice to be enough for all physical
situations, the next question is what should that choice be. \ It is
important to stress that the advanced/retarded Green functions involve
respectively a future or a past position measured at \emph{another spatial
point} by another clock synchronized a la Einstein, which is not
contradictory to causality like it would be using the Newtonian future in a
Newtonian mechanics, for example. In the theory of relativity one is not
allowed to compare times that are not measured at the same point. To falsify
the "known" future of the other spatial point, one would have to travel to
this spatial point. By the definition of the light-cone, travelling at the
speed of light should arrive precisely when the predicted future happens, so
that no falsification of the "known future" is possible. The remaining
rationale for the most popular choice (retarded-only Green function) are
based on the analysis of a single charge acted upon by non-electromagnetic
forces that start suddenly at $t=0$. Combined with the further assumption
that acceleration is the cause of the far-fields, this popular argument
leads to the retarded-only Green function as the unique choice where the
far-fields vanish for $t<0$. This tendentious argument and the one-body
picture behind it overlook the possibility of \emph{bounded motions }of two-
or more- charges\emph{\ }supported by \emph{their} \emph{own}
electromagnetic forces, as the Schoenberg-Schild orbits for example\cite%
{Schild};---For such globally-bounded motions the charges have accelerated
all the way back in the infinite past \emph{and} the electromagnetic
far-fields exist for \emph{all times} regardless of the choice of the Green
function. For example, along a circular motion with a small velocity the
future field and the past field approximately coincide for an increasing
series of distances approximately proportional to multiples of the rotation
period, so that the retarded Green function and the advanced Green function
predict very similar field-patterns anyway. Moreover the equations of motion
obtained using the advanced Green function involve the advanced acceleration
of the other particle, which can be solved for this most-advanced
acceleration yielding delay-only equations of motion, Eq. (22) of Ref. \cite%
{JMP}, a construction that needs no future information to define solutions.
We are of the opinion that the study of many-body electromagnetic-only
problems should guide the choice of \ the Green%
\'{}%
s function, and not vice-versa. For that the knowledge of the possible
bounded motions of a finite number of charges is essential\cite{martinez}.
Here we have shown that bounded two-body orbits are impossible in the
Lorentz-Dirac theory, so that interesting dynamics necessitates three bodies
at the best in the Dirac theory. The inexistence of unperturbed
globally-bounded two-body motions is a shortcoming of the Dirac theory if
one is looking for a sensible dynamical system to describe atomic physics
classically. On the other hand, there are several selling points for the
action-at-a-distance theory: (i) a known one-parameter family of bounded
circular-orbits for two-body motion\cite{Schild}, (ii) the regularity of the
point-charge limit and the absence of self-interaction and runaways\cite%
{EliezerReview}, (iii) The ill-posedness of the backward equations of motion
in any electrodynamics of point charges but the action-at-a-distance theory,
i.e., the backward equations of motion define the acceleration as a function
of its derivative and second derivative in the past, Eq. (25) of Ref. \cite%
{JMP}. As discussed in Ref. \cite{JMP}, along a backward integration one
constructs an acceleration that is only continuous until the first breaking
point, but later on its first and second derivatives in the past are needed.
Since these extra derivatives exist only if the third and fourth derivatives
of the initial history existed, recursively one should have started with a $%
C^{\infty }$ initial segment. This recursion makes it impossible to start
with generic data, so that if a solution ever existed one should start a
backward integration from very special $C^{\infty }$ data, which would be
spoiled by the numerical calculations. Moreover, it turns out that backward
integration\emph{\ would be} absolutely necessary because the forward
equations have explosive runaway solutions\cite{JMP}, so that something is
not well for the Lorentz-Dirac theory. On the contrary, action-at-a-distance
electrodynamics yields neutral-delay backward equations for two-body motion,
rather than ill-posed, and last (v)\ Inside the action-at-a-distance theory,
bounded universes are not required to satisfy the A.H., even though the A.H.
can provide an improved stability to counter energy losses, as follows;---If
only the energy flux of the semi-sum field vanishes, an offending
perturbation of size $\varepsilon $ in the retarded far-field (or in the
advanced far-field) perturbs the energy flux Eq. (\ref{PoyntingX}) at $%
O(\varepsilon )$, while if the A.H. holds the perturbed flux is $%
O(\varepsilon ^{2})$ by Eq. (\ref{PoyntingX}). Therefore the A.H. can be
thought as a stronger boundary condition for stable universal dynamics in
the action-at-a-distance theory, while in the Dirac theory the A.H. is
necessary for a vanishing energy flux.


\bigskip

\bigskip

\begin{references}

\bibitem{Fey-Whe} J. A. Wheeler and R. P. Feynman {\it Rev.Mod. Phys.} {\bf 17}, 
157 (1945);  J. A. Wheeler and R. P. Feynman {\bf 21}, 425 (1949). 

\bibitem{Dirac} P. A. M.Dirac, {\it Proceedings of the Royal Society of London, ser. A}
{\bf 167},148 (1938).

\bibitem{Narlikar}F. Hoyle and J. V. Narlikar, {\em Lectures on Cosmology and Action at a Distance Electrodynamics }, 
World Scientific, London (1996).

\bibitem{Nodvik}J.S.Nodvik, {\it Ann. Phys. (N.Y.)} {\bf 28} 225 (1964).

\bibitem{EliezerReview}C. Jayaratnam Eliezer, {\it Reviews of Modern Physics} {\bf 19} (1947). 

\bibitem{Jackson}J.D. Jackson, {\em Classical Electrodynamics} Second Edition, 
John Wiley and Sons, New York(1975) ( Eq. 6.61 of page 224).

\bibitem{Schoenberg}M. Schoenberg, {\it Physical Review} {\bf 69} 211 (1946).

\bibitem{Schild}A. Schild, {\it Physical Review} {\bf 131}, 2762 (1963).

\bibitem{Starus}A. Staruszkiewicz, {\it Acta Physica Polonica B} {\bf 33} 2041 (2002).


\bibitem{radiationless}G.H.Goedecke, {\it Physical Review} {\bf 135} B281 (1964) 
and J. B. Arnett and G. H. Goedecke, {\it Physical Review} {\bf 168} 1424 (1968).


\bibitem{stiff-hydrogen} J. De Luca, { \it Physical Review E } {\bf 73}, 026221 (2006).

\bibitem{Boyer}T. Boyer, {\it Physical Review D} {\bf 11}, 790 (1975).


\bibitem{dissipativeFokker} J. De Luca, { \it Physical Review E } {\bf 71}, 056210 (2005).

\bibitem{JMP}J. De Luca, {\it J. Math. Phys.} {\bf 48}, 012702 (2007).  

\bibitem{EFY2}E.B.Hollander and J. De Luca, {\it Chaos} {\bf 14} 1093 (2004).


\bibitem{martinez}D. J. Louis-Martinez, {\it Phys. Lett. A } {\bf 320} 103 (2003).





\bibitem{Eliezertheorem}C. J. Eliezer, {\it Proc. Cambridge Philos. Soc. } {\bf 39} 173 (1943).





\end{references}
\end{document}